\documentclass[prc,aps,nofootinbib,twocolumn]{revtex4}

\usepackage{epsfig}
\usepackage{color}
\usepackage[latin1]{inputenc}
\usepackage{float,amsmath}
\usepackage{graphicx}

\begin{document}

\title{MARTINI: An event generator for relativistic heavy-ion collisions}

\author{Bj\"orn Schenke}
\affiliation{Department of Physics, McGill University, Montreal, Quebec, H3A\,2T8, Canada}

\author{Charles Gale}
\affiliation{Department of Physics, McGill University, Montreal, Quebec, H3A\,2T8, Canada}

\author{Sangyong Jeon}
\affiliation{Department of Physics, McGill University, Montreal, Quebec, H3A\,2T8, Canada}

\begin{abstract}
\hyphenation{MARTINI}
We introduce the Modular Algorithm for Relativistic Treatment of heavy IoN Interactions (MARTINI),
a comprehensive event generator for the hard and penetrating probes in high energy
nucleus-nucleus collisions.
Its main components are a time evolution model for the soft background, PYTHIA~8.1
and the McGill-AMY parton evolution scheme including radiative as well as elastic processes. 
This allows us to generate full event configurations in the high $p_T$
region that take into account thermal QCD and QED effects as well as effects of the evolving medium.
We present results for the neutral pion nuclear modification factor in Au+Au collisions at RHIC
as a function of $p_T$ for different centralities, and also as a function of the angle with respect to the reaction plane for
non-central collisions.
Furthermore, we study the production of high transverse momentum photons incorporating a complete set of photon-production channels.
\end{abstract}

\maketitle


\section{Introduction}
High transverse momentum jets emerging from the central rapidity region
in heavy-ion collisions provide important information on the produced hot quark-gluon plasma (QGP).
To extract this information from the experimental data, it is important to develop a good theoretical understanding
of the interactions of hard partons with the medium.
Therefore the energy loss of hard partons traversing the medium has been under extensive theoretical investigation.
Gluon bremsstrahlung including the Landau-Pomeranchuk-Migdal (LPM) \cite{Migdal:1956tc} effect 
has been proposed as the dominant mechanism for energy loss and
different theoretical formalisms have been developed and applied to describe it
 \cite{Baier:1996kr, Kovner:2003zj, Zakharov:1996fv, Gyulassy:2000er, Wang:2001ifa, 
Zhang:2003yn, Majumder:2004pt, Majumder:2007hx, Arnold:2001ms, Arnold:2001ba, Arnold:2002ja}. 
Also, binary elastic scattering off thermal partons is potentially important for 
the energy loss and momentum broadening of high-$p_T$ partons.
In \cite{Qin:2007rn}, elastic energy loss was combined with the AMY 
\cite{Arnold:2001ms, Arnold:2001ba, Arnold:2002ja} radiative energy loss within the McGill-AMY evolution formalism, 
and in \cite{Schenke:2009ik} the description of the collisional processes was further improved.

The main goal of these and other calculations is to create a quantitative basis for
the ``tomography'' of the hot and dense nuclear medium created in heavy-ion collisions.
Among the ``tomographic variables'' is the nuclear modification factor $R_{AA}$, 
which is defined as the ratio of the hadron yield in A+A collisions to that in
binary-scaled p+p interactions. 
A variety of computations that differ 
significantly in the applied energy loss mechanism can reproduce the measured 
$R_{AA}$ in central Au+Au collisions at $\sqrt{s}=200~{\rm AGeV}$ 
at the Relativistic Heavy-Ion Collider (RHIC), given the present experimental errors (see \cite{Bass:2008rv} 
for a systematic comparison of three different formalisms).
For this reason the tomographic usefulness of $R_{AA}$ in central collisions has become questionable \cite{Renk:2006qg}. 
More differential observables, such as $R_{AA}$ as a function of both $p_T$ and azimuth in non-central collisions
\cite{Majumder:2006we,Adler:2006bw,Wei:2009mj,Afanasiev:2009iv}, can help to improve this situation.

Other important observables in heavy-ion collisions are electromagnetic probes, such as photons and dileptons.
Because of the small electromagnetic coupling, once produced, they usually escape the medium without further interaction
and thus provide undistorted information on the early stages of a heavy-ion collision.  
Photon production from nuclear collisions at RHIC has been calculated
in \cite{Turbide:2005fk,Turbide:2007mi,Qin:2009bk}
using 1+1 dimensional Bjorken evolution, 2+1 dimensional, or 3+1 dimensional relativistic hydrodynamic 
evolution for the background, respectively. 
Photons in nuclear collisions come from a variety of sources, namely
direct photons, fragmentation photons, jet-plasma photons and thermal photons. 
Direct photons are predominantly produced from hard collisions in the early stage of the heavy-ion collision via
annihilation and Compton scattering processes. Fragmentation photons are produced from the surviving high energy
jets after their passing through the thermal medium. 
Thermal photons have a negligible contribution at high $p_T$ and thus are not relevant in this range.
On the other hand, jet-plasma photons from photon radiation $q\rightarrow q\gamma$ (bremsstrahlung photons) and 
jet-photon conversion via Compton and annihilation processes
have been shown to be very important for the understanding of experimental data for photon production
in Au+Au collisions at RHIC \cite{Fries:2002kt,Turbide:2005fk,Turbide:2007mi,Qin:2009bk}.

For a best possible comparison of the theoretical description with experimental data,
we incorporate the McGill-AMY formalism \cite{Arnold:2001ms, Arnold:2001ba, Arnold:2002ja,Qin:2007zz,Qin:2007rn} 
for radiative energy loss as well 
as elastic processes \cite{Schenke:2009ik} into a new event generator, dubbed
Modular Algorithm for Relativistic Treatment of heavy IoN Interactions (MARTINI). 
Its main ingredients are PYTHIA~8.1 \cite{Sjostrand:2007gs,Sjostrand:2006za} to generate the hard partons from the 
individual nucleon-nucleon collisions and to take care of the final fragmentation into hadrons,
the evolution of the background medium (e.g. from  hydrodynamic models), and the McGill-AMY parton evolution formalism.
This way it is possible to study hard observables in heavy-ion collisions theoretically on an event-by-event basis,
keeping information on correlations.

Several Monte Carlo simulations for heavy-ion collisions have been or are being developed 
\cite{Wang:1991hta,Dainese:2004te,Lokhtin:2005px,Lokhtin:2008xi,Wicks:2008ta,Renk:2008pp,Renk:2008zr,
Zapp:2008gi,Armesto:2009fj}. The implementation of medium effects and use of approximations varies significantly 
between the different models. 
MARTINI is the first Monte Carlo simulation to include AMY bremsstrahlung
combined with elastic processes and in-medium photon production.
It is very flexible due to its ability to incorporate any soft background evolution that provides information
on the temperature and flow velocities of the medium.
Currently, hydrodynamic evolution calculations from four different groups 
\cite{Eskola:2005ue,Kolb:2000sd,Kolb:2002ve,Kolb:2003dz,Nonaka:2006yn,Heinz:2009cv} have been implemented. 

In this work we present first results on hard observables obtained with MARTINI. These include the azimuth averaged 
$R_{AA}$ as a function of $p_T$, $R_{AA}$ as a function of the azimuth and $p_T$, as well as photon yields and
photon $R_{AA}$.

This paper is organized as follows.
We review the McGill-AMY evolution formalism and present transition rates for radiative and elastic processes in Section \ref{mcgill-amy}.
The Monte Carlo simulation is introduced in Section \ref{martini}, results are presented in Section \ref{results}, 
and we conclude and present an outlook in Section \ref{conclusions}.


\section{The McGill-AMY evolution formalism}
\label{mcgill-amy}
At the core of MARTINI lies the McGill-AMY formalism for jet evolution in a dynamical thermal medium.
Here, we briefly review this formalism and discuss its implementation into MARTINI in the following section.

In previous works \cite{Qin:2007zz, Qin:2007rn, Qin:2009bk} the evolution of the jet momentum
distribution in a hydrodynamic background was computed within the McGill-AMY scheme.
This evolution is governed by a set of coupled Fokker-Planck type rate equations of the 
form
\begin{align}\label{jet-evolution-eq}
\frac{dP(p)}{dt}\!=\!\int_{-\infty}^{\infty}\!\!\!\!\!\!dk
\left(\!P(p{+}k) \frac{d\Gamma(p{+}k,k)}{dk} - P(p)\frac{d\Gamma(p,k)}{dk}\!\right)\,,
\end{align}
where ${d\Gamma(p,k)}/{dk}$ is the transition rate for
processes where partons of energy $p$ lose energy $k$. 
The $k<0$ part of the integration incorporates processes which increase
a parton's energy. 

In the AMY finite temperature field theory approach,
the energy loss of hard partons is considered in a medium at an asymptotically
high temperature so that the QCD coupling is weak.
In this regime, an analytical treatment is possible, 
and a hierarchy of parametrically separated scales $T \gg gT \gg g^2T$ emerges.
This allows radiative transition rates to be calculated by means of integral equations
\cite{Arnold:2002ja}, which correctly reproduce both the Bethe-Heitler and the LPM results
in their respective limits. 
They are given by \cite{Jeon:2003gi, Turbide:2005fk}:
\begin{eqnarray}\label{eq:dGamma}
\frac{d\Gamma}{dk}(p,k) & = & \frac{C_s g^2}{16\pi p^7}
        \frac{1}{1 \pm e^{-k/T}} \frac{1}{1 \pm e^{-(p-k)/T}}
\nonumber \\ && \times \left\{ \begin{array}{cc}
        \frac{1+(1{-}x)^2}{x^3(1{-}x)^2} & q \rightarrow qg \\
        N_{\rm f} \frac{x^2+(1{-}x)^2}{x^2(1{-}x)^2} & g \rightarrow q\bar{q} \\
        \frac{1+x^4+(1{-}x)^4}{x^3(1{-}x)^3} & g \rightarrow gg \\
        \end{array} \right\}
\nonumber \\ && \times \int \frac{d^2 \mathbf{h}}{(2\pi)^2} 2 \mathbf{h} \cdot {\rm Re}\: \mathbf{F}(\mathbf{h},p,k) \,,
\end{eqnarray}
where $g$ is the strong coupling constant ($\alpha_{\rm s}=g^2/(4\pi)$), $N_{\rm f}$ is the number of flavors,
$C_s$ is the quadratic Casimir relevant for the process, and $x\equiv k/p$ is the
momentum fraction of the radiated
gluon (or the quark, for the case $g \rightarrow q\bar{q}$). 
$\mathbf{h} \equiv (\mathbf{k} \times \mathbf{p}) \times \mathbf{e}_\parallel$ determines
how non-collinear the final state is, where $\mathbf{e}_\parallel$ is the unit vector in a direction nearly collinear 
with $\mathbf{k}$, $\mathbf{p}$ and $\mathbf{p}+\mathbf{k}$ that can be fixed by convention (see \cite{Arnold:2002ja}). 
Parametrically $\mathbf{h}$ is of order $gT^2$ and hence small compared
to $\mathbf{p}\cdot \mathbf{k}$. Therefore $\mathbf{h}$ can be taken as a two-dimensional vector in transverse space.
$\mathbf{F}(\mathbf{h},p,k)$ is the solution of the following integral equation~\cite{Jeon:2003gi, Turbide:2005fk}:
\begin{eqnarray}\label{eq:integral_eq1}
2\mathbf{h} &=&
        i \delta E(\mathbf{h},p,k) \mathbf{F}(\mathbf{h}) + g_s^2 \int \frac{d^2 \mathbf{q}_\perp}{(2\pi)^2}
C(\mathbf{q}_\perp) \nonumber \\ && \times
   \Big\{ (C_s-C_{\rm A}/2)[\mathbf{F}(\mathbf{h})-\mathbf{F}(\mathbf{h}{-}k\,\mathbf{q}_\perp)]
        \nonumber \\ &&
        + (C_{\rm A}/2)[\mathbf{F}(\mathbf{h})-\mathbf{F}(\mathbf{h}{+}p\,\mathbf{q}_\perp)]
        \nonumber \\ &&
        +(C_{\rm A}/2)[\mathbf{F}(\mathbf{h})-\mathbf{F}(\mathbf{h}{-}(p{-}k)\,\mathbf{q}_\perp)] \Big\} \, .
\end{eqnarray}
Here $\delta E(\mathbf{h},p,k)$ is the energy difference between the final and the initial states:
\begin{eqnarray}\label{eq:integral_eq2}
\delta E(\mathbf{h},p,k) &=& \frac{\mathbf{h}^2}{2pk(p{-}k)} + \frac{m_k^2}{2k} + \frac{m_{p{-}k}^2}{2(p{-}k)} -
\frac{m_p^2}{2p}\, , \ \ \ \ \ \
\end{eqnarray}
and $m^2$ are the medium induced thermal masses. $C(\mathbf{q}_\perp)$ is the differential rate to exchange
transverse (to the $\parallel$ direction) momentum $\mathbf{q}_\perp$. In a hot thermal medium, its value at leading order in
$\alpha_{\rm s}$ is \cite{Aurenche:2002pd}
\begin{eqnarray}\label{eq:Cq}
C(\mathbf{q}_\perp) = \frac{m_D^2}{\mathbf{q}_\perp^2(\mathbf{q}_\perp^2{+}m_D^2)} \, , \ \ \ m_D^2 = \frac{g_s^2 T^2}{6}
(2 N_{\rm c} {+} N_{\rm f})\, . \ \ \
\end{eqnarray}
For the case of $g\rightarrow q\bar{q}$, $(C_s-C_{\rm A}/2)$ should appear as the prefactor on the term
containing $\mathbf{F}(\mathbf{h}-p\,\mathbf{q}_\perp)$ rather than $\mathbf{F}(\mathbf{h}-k \,\mathbf{q}_\perp)$.

Transition rates for elastic processes can also be computed in perturbative QCD. They are given by \cite{Schenke:2009ik}
\begin{align}\label{eq:tr}
 \frac{d\Gamma}{d\omega}~(p,\omega,T)=&d_q\int \frac{d^3q}{(2\pi)^3}\int \frac{d^3q^\prime}{(2\pi)^3}\frac{2\pi}{16pp^\prime q q^\prime}\notag\\ 
 &\times\delta(p-p^\prime-\omega)\delta(q^\prime-q-\omega)\notag\\
 &\times|\mathcal{M}|^2 f(q,T)(1\pm f(q^\prime,T))\,,
\end{align}
where $p=|\mathbf{p}|$ and $p^\prime=|\mathbf{p}^\prime|$ are the absolute values of the three-momenta of the incoming and outgoing hard parton, respectively, and $q=|\mathbf{q}|$ and $q^\prime=|\mathbf{q}^\prime|$ are those of the incoming and outgoing thermal parton.
 $\omega=p-p^\prime=q^\prime-q$ is the transferred energy.
The distribution functions $f$ are either Fermi-Dirac or Bose-Einstein distributions depending on the 
nature of the thermal parton involved. The $+$ or $-$ sign appears accordingly, 
with $-$ for Pauli blocking and $+$ for Bose enhancement, and $d_q$ describes the degeneracy of the thermal parton.
Integrations over $q$ and $q^\prime$ in Eq.~(\ref{eq:tr}) can be rewritten in terms of the transferred momentum 
$k$. Then, employing an appropriate separation of the regimes of soft and hard momentum exchange,
and doing a hard thermal loop resummation in the soft regime to cure the infrared divergence allows for a numerical 
evaluation of Eq.~(\ref{eq:tr}).
It is even possible to extend the expression for the soft regime to all momenta \cite{Djordjevic:2006tw,Schenke:2009ik}, which leads
to a good approximation of the total result and avoids the introduction of an intermediate scale.
The transition rate as a function of both the transferred energy and momentum $d\Gamma/d\omega dk$ 
is obtained from Eq.~(\ref{eq:tr}) after rewriting and omitting the integration over the transferred momentum $k$.
Using $d\Gamma/d\omega dk$ will allow us to sample both the transferred energy and the transferred three-momentum in an elastic process.

Both radiative and elastic transition rates were computed
and tabulated as functions of the parton energy $p$, as well as the radiated energy $k$ or transferred energy $\omega$ and momentum $k$, 
respectively, and are sampled within MARTINI. Note that since the radiated partons are on-shell $\omega=k$, while the transferred
parton in an elastic collision can be far off-shell, so that $\omega$ and $k$ are treated separately.
As opposed to the radiative transition rates, the dependence on the coupling of the elastic rates is non-trivial and the rates are 
also tabulated as functions of the coupling $\alpha_{\rm s}$.

Apart from the processes described above we include gluon-quark and quark-gluon conversion
due to Compton and annihilation processes, as well as the QED processes of
photon radiation $q\rightarrow q\gamma$ and jet-photon conversion.
The transition rate for the photon radiation process is calculated analogously to the gluon radiation 
\cite{Arnold:2001ms,Arnold:2002ja}, and in the limit $E\gg T$ the transition rate for quark-photon conversion is given by
\begin{eqnarray}\label{eq:conversion}
  \frac{d\Gamma^{\rm conv}_{q\rightarrow \gamma}}{d\omega}(E,\omega) &=&\nonumber\\
  &&\hspace{-2cm}\left(\frac{e_{f}}{e}\right)^2
  \frac{2\pi \alpha_{\rm e} \alpha_{\rm s} T^2}{3 E} \left(\frac{1}{2}{\rm ln}\frac{ET}{m_q^2}-0.36149\right)\delta(\omega)\,,
\end{eqnarray}
where $m_q$ is the thermal quark mass, $m_q^2=g^2T^2/6$,
$e_{f}$ is the charge of a quark with flavor $f$, 
$\alpha_{\rm e}$ is the fine structure constant, 
and the delta function indicates that we neglect the energy loss of the quark during the process of conversion.
The conversion rate $d\Gamma^{\rm conv}_{q\rightarrow g}/d\omega$ is equal to that in 
Eq.~(\ref{eq:conversion}) times a color factor $C_F=4/3$, and 
\begin{equation}
\frac{d\Gamma^{\rm conv}_{g\rightarrow q}}{d\omega} =
 N_{\rm f} \frac{N_c}{N_c^2 - 1} \frac{d\Gamma^{\rm conv}_{q\rightarrow g}}{d\omega}\,,
\end{equation}
which follows from interchanging the 
initial with the final jet.

\section{Monte Carlo simulation}
\label{martini}
MARTINI solves the rate equations (\ref{jet-evolution-eq}) using Monte Carlo methods.
Instead of evolving a probability distribution
$P(p)$, every event and every hard parton is treated individually. 
This way we obtain information on the full microscopic event configuration in the high momentum regime, including correlations,
which allows for a very detailed analysis and offers a direct interface between theory and experiment.
The average over a large number of events will correspond to the solution found by solving (\ref{jet-evolution-eq}) 
for the probability distribution. 

We follow the evolution of one event to describe the basic functionality of the simulation.
First, the number of individual nucleon-nucleon collisions that produce partons with a certain minimal
transverse momentum $p_T^{\rm min}$ is determined from the total inelastic cross-section,
provided by PYTHIA. 
The initial transverse positions $\mathbf{r}_\perp$ of these collisions
are determined by the initial jet density distribution $\mathcal{P}_{AB}(b,\mathbf{r}_\perp)$, which for
A+B collisions with impact parameter $\mathbf{b}$ is given by
\begin{eqnarray}
\mathcal{P}_{AB}(b,\mathbf{r}_\perp) &=& \frac{T_A(\mathbf{r}_\perp + \mathbf{b}/2)T_B(\mathbf{r}_\perp -
\mathbf{b}/2)}{T_{AB}(b)}.
\end{eqnarray}
Here we use a Woods-Saxon form for the nuclear density function,
$\rho(\mathbf{r}_\perp,z)={\rho_0}/[{1+\exp(\frac{r-R}{d})}]$, to evaluate the nuclear thickness function
$T_A(\mathbf{r}_\perp)=\int dz \rho_A(\mathbf{r}_\perp,z)$ and the overlap function of two nuclei
$T_{AB}(b)=\int d^2r_\perp T_A(\mathbf{r}_\perp) T_B(\mathbf{r}_\perp+\mathbf{b})$. 
The values of the parameters $R=6.38~{\rm fm}$ and $d=0.535~{\rm fm}$ are taken from \cite{DeJager:1974dg}.
The initial parton distribution functions can be selected with the help of the Les Houches Accord PDF Interface (LHAPDF) 
\cite{Whalley:2005nh}.
We include nuclear effects on the parton distribution functions 
using the EKS98 \cite{Eskola:1998df} or EPS08 \cite{Eskola:2008ca} parametrization, by user choice.
In practise we sample the initial positions of nucleons in nucleus A and B, superimpose the transverse 
areas depending on the impact parameter and then divide the overlap region into
circles with area $\sigma_{\rm inel}$. In three dimensions these are tubes and we determine how many jet events
with given $p_T^{\rm min}$ occur within such a tube using the number of nucleons from A and B in the tube, and the
probability for a jet event $\sigma_{\rm jet}(p_T^{\rm min})/\sigma_{\rm inel}$ for each of their combinations.

The soft medium is described by hydrodynamics or other models, which provide information on the system's local temperature 
and flow velocity $\boldsymbol{\beta}$. 
Before this medium has formed, i.e., before the hydrodynamic evolution begins ($\tau<\tau_0$),
the partons shower as in vacuum. Currently we include the complete vacuum shower, because there is no apparent reason why the vacuum splittings should
end immediately once the medium has formed. Since most of the vacuum shower occurs before the medium has formed, this is a reasonable approximation. 
We have also tested the other extreme case where we stop the vacuum evolution at the virtuality scale $Q_{\rm min}=\sqrt{p_T/\tau_0}$, determined by the
time $\tau_0$ at which the medium evolution begins. This modifies the parton distribution such that the strong coupling constant has to be
chosen approximately $10\%$ larger to describe the pion $R_{AA}$.
At this stage, we do not include final state radiation (FSR) of the partons that
have left the medium - all vacuum showers take place before the medium evolution and there is no interference 
between the medium and vacuum showers. The improvement of this point is the subject of future work. 
Once $\tau>\tau_0$ for a certain parton, its in-medium evolution begins.
The partons move through the background according to their velocity.
To compute probabilities for any one of the above described in-medium processes, 
we perform a Lorentz-boost into the rest-frame of the fluid cell at the parton's position and 
determine the transition rates according to the local temperature and the parton's energy in this local rest-frame.
The probability for a parton to undergo any process during a time step of length $\Delta t$ is given by
$\Delta t\, \Gamma(p,T)_{\rm total}$, where $\Gamma(p,T)_{\rm total}=\int dk d\Gamma_{\rm total}/dk$ for the radiative processes, 
and the integral over both $\omega$ and $k$ for the elastic processes. 
$d\Gamma_{\rm total}/dk$ is the sum over all possible transition rates, which include those for photon production. 

In case that some process occurs, we decide which one it is according to the relative weights of the different
processes at the given temperature and parton energy. 
We sample the radiated or transferred energy from the transition rate of the occurring process using 
the rejection method. In case of an elastic process, we also sample the transferred transverse momentum, while
for radiative processes we assume collinear emission for now, which is a good approximation in the weak coupling limit
since the emission angle is of order $g$ \cite{Arnold:2002ja}.
After energy and momentum are transferred, we boost back into the laboratory frame, where the parton continues to move along its trajectory.
Radiated partons are also further evolved if their momentum is above a certain threshold $p_{\rm min}$, which
can be set (typically we choose $p_{\rm min}\simeq 3\,{\rm GeV}$). This leads to a growing and broadening in-medium shower.
The overall evolution of a parton stops once its energy in a fluid cell's rest frame falls below the limit
of $4T$, where $T$ is the local temperature. As both radiative and elastic transition rates were computed in the limit
$E\gg T$, it would not make sense to attempt to learn anything about the evolution of partons with $E\simeq T$ within this formalism.
For partons that stay above that threshold, the evolution ends once they enter the hadronic phase of the background medium.
In the mixed phase, processes occur only for the QGP fraction.

\begin{figure}[htb]
  \begin{center}
    \includegraphics[width=8.5cm]{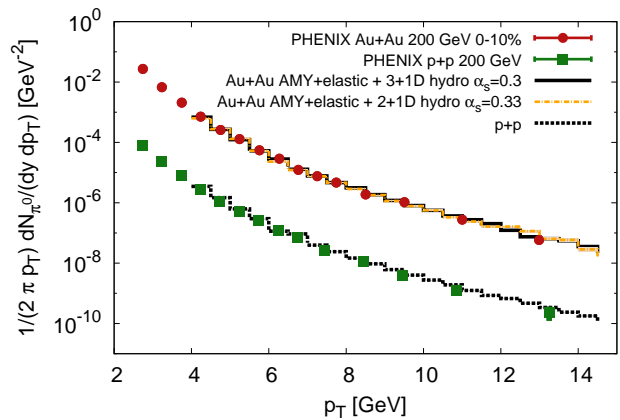}
    \caption{(Color online) Neutral pion spectrum in p+p and $0-10\%$ central Au+Au collisions at RHIC energy 
      compared to data by PHENIX \cite{pp:2005,Adler:2006bw}.
      }
    \label{fig:pi0-pp}
  \end{center}
\end{figure}
When all partons have left the QGP phase, hadronization is performed by PYTHIA, to which the complete
information on all final partons is passed. Because PYTHIA uses the Lund string fragmentation model 
\cite{Andersson:1983ia,Sjostrand:1984ic}, it is essential to
keep track of all color strings during the in-medium evolution. This includes generating new strings when a gluon is emitted or
during a conversion process. In the latter case, new string ends have to be attached to thermal partons, 
whose momenta are sampled from Fermi or Bose distributions.
Being provided with such consistent information on the color string structure, PYTHIA is able to perform the fragmentation.
This concludes the evolution of one event.

The concept of MARTINI is modular, such that we can turn on and off different processes independently, and use different hydrodynamic
or other data inputs. In principle it is also possible to extend MARTINI to use different energy loss formalisms. 
The parameters are set using the same interface as PYTHIA and all options for PYTHIA can still be modified by the user.

\section{Results}
\label{results}


\subsection{Pion production and nuclear modification factor $\pi^0 R_{AA}(p_T)$}
We begin by showing the MARTINI results for the spectrum of neutral pions in p+p collisions at RHIC 
energy ($\sqrt{s}=200\,{\rm GeV}$)
as well as in central (0-10\%) Au+Au collisions compared to data by PHENIX \cite{pp:2005,Adler:2006bw} in Fig.~\ref{fig:pi0-pp}.
The calculations were performed using CTEQ5L parton distribution functions \cite{Lai:1999wy} 
including nuclear shadowing effects using the EKS98 parametrization \cite{Eskola:1998df}.
In the shown results for pion production, isospin effects were ignored. 
We checked that they have no big effect (less than 5\%), however, we included them 
in calculations of photon production where they become important (15-20\%).
Au+Au calculations take into account both radiative and elastic processes in the medium described by 
either the 2+1 dimensional hydrodynamics of \cite{Kolb:2000sd,Kolb:2002ve,Kolb:2003dz} 
or the 3+1 dimensional hydrodynamics of \cite{Nonaka:2006yn}, using a coupling constant $\alpha_{\rm s}=0.33$ or
$\alpha_{\rm s}=0.3$, respectively. $\alpha_{\rm s}$ 
was adjusted to describe the experimental measurement of the neutral pion nuclear modification factor
$R_{AA}$ in most central collisions (see below). For both hydro evolutions $\tau_0=0.6\,{\rm fm}/c$.
PYTHIA parameters were tuned to fit the p+p data, however there is still freedom and possibly an even better set of parameters 
can be found. Using the same PYTHIA settings, we find very good agreement for both p+p and Au+Au spectra.
Note that in order to achieve higher statistics we used certain cutoffs for the minimal $p_T$ produced in a nucleon-nucleon 
collision in PYTHIA for the Au+Au calculation. The shown result is a combination of runs with minimal 
$p_T^{\rm min}=1.5\,{\rm GeV}$ for $p_T\leq 5.5\,{\rm GeV}$, $p_T^{\rm min}=2\,{\rm GeV}$ for 
$5.5\,{\rm GeV}\leq p_T \leq 6.5\,{\rm GeV}$, and $p_T^{\rm min}=4\,{\rm GeV}$ above that. 
Note that the shown results are averages over typically $\sim5\cdot\,10^8$ simulated events.
\begin{figure}[tb]
  \begin{center}
    \includegraphics[width=8.5cm]{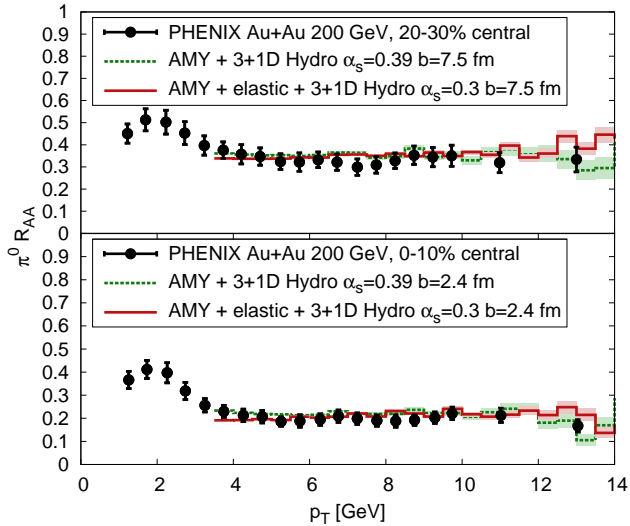}
    \caption{(Color online) The neutral pion nuclear modification factor for mid-central (upper panel) 
      and central (lower panel) collisions at RHIC with $\sqrt{s}=200\,{\rm GeV}$.
      MARTINI results ($b=2.4\,{\rm fm}$ and $b=7.5\,{\rm fm}$) using 3+1 dimensional hydro evolution \cite{Nonaka:2006yn}
      and only radiative processes (dashed lines) 
      and both radiative and elastic processes (solid lines) with different $\alpha_{\rm s}$ 
      compared to PHENIX data from \cite{Adare:2008qa}.}
    \label{fig:raa}
  \end{center}
\end{figure}

\begin{figure}[tb]
  \begin{center}
    \includegraphics[width=8.5cm]{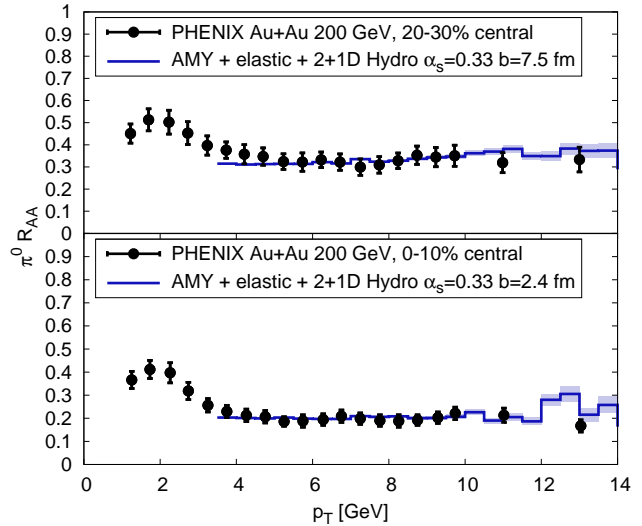} 
    \caption{(Color online) The neutral pion nuclear modification factor for mid-central (upper panel) 
      and central (lower panel) collisions at RHIC with $\sqrt{s}=200\,{\rm GeV}$.
      MARTINI results ($b=2.4\,{\rm fm}$ and $b=7.5\,{\rm fm}$) using 2+1 dimensional hydro evolution 
      \cite{Kolb:2000sd,Kolb:2002ve,Kolb:2003dz}
      and both radiative and elastic processes compared to PHENIX data from \cite{Adare:2008qa}.}
    \label{fig:kolbraa}
  \end{center}
\end{figure}

In Fig.~\ref{fig:raa} we present the results for the neutral pion nuclear modification factor, defined by
\begin{equation}
  R_{AA}=\frac{1}{N_{\rm coll}(b)} \frac{dN_{AA}(b)/d^2p_T dy}{dN_{pp}/d^2p_Tdy}\,,
\end{equation}
in Au+Au collisions at RHIC measured at mid-rapidity in two different centrality classes (0-10\%) and (20-30\%),
employing the corresponding average impact parameters, $2.4\,{\rm fm}$ and $7.5\,{\rm  fm}$.
We compare with the experimental measurements by PHENIX \cite{Adare:2008qa}.
All parameters are the same as in the calculation shown in Fig.~\ref{fig:pi0-pp}, i.e.,
hydrodynamic background evolution from \cite{Nonaka:2006yn} and $\alpha_{\rm s} = 0.3$.
The same value of $\alpha_{\rm s}$ is used in all following calculations. 
(There are no additional parameters for the later calculation of 
the azimuthal dependence of $R_{AA}$ or the high-$p_T$ photon production.)
We find very good agreement with the data for both centrality classes.

The value for $\alpha_{\rm s}$ for which the data is described if we only include radiative processes
is $\alpha_{\rm s}\simeq 0.39$, higher than for the calculations in \cite{Qin:2007rn},
where it was $0.33$. 
Differences in the treatment of vacuum showers and the fragmentation scheme are responsible for this difference.
We have observed that when varying parameters in PYTHIA, such as the factorization or renormalization scale, or
parameters in the fragmentation function, we need slighty different $\alpha_s$ to describe the data.
This freedom leads to an approximately $10$ - $20\,\%$ uncertainty in $\alpha_s$.
The value of $\alpha_{\rm s}$ is again closer to that used in \cite{Qin:2007rn} ($\alpha_s=0.27$) when including elastic processes, 
because the elastic energy loss in our formulation is slightly larger than that implemented in \cite{Qin:2007rn} 
- see \cite{Schenke:2009ik} for details on this.
Like previous works (e.g. \cite{Qin:2007rn}) we find that elastic processes are important for the computation of $R_{AA}$.

Importantly, our approach has the potential to reveal the effect of different soft background models
on hard observables. Fig.~\ref{fig:kolbraa} shows the result for $R_{AA}$ using 
the 2+1 dimensional hydrodynamic evolution from \cite{Kolb:2000sd,Kolb:2002ve,Kolb:2003dz}. As initial temperatures
are typically lower than in the 3+1 dimensional hydro evolution, we need to increase $\alpha_{\rm s}$ to $0.33$ to describe the
neutral pion $R_{AA}$. Further effects are demonstrated in the following section.


\subsection{Angular dependence of the nuclear modification factor $\pi^0 R_{AA}(p_T,\Delta\phi)$}
\begin{figure}[htb]
  \begin{center}
    \includegraphics[width=4cm]{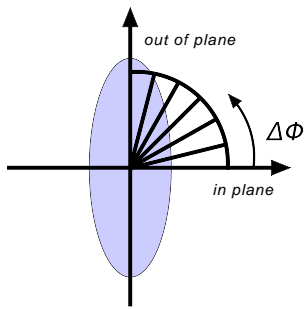}
    \caption{(Color online) Binning in azimuth $\Delta \phi$ for the calculation of $R_{AA}(p_T, \Delta\phi)$ in non-central collisions.}
    \label{fig:anglephi}
  \end{center}
\end{figure}

\begin{figure}[htb]
  \begin{center}
    \includegraphics[width=8.5cm]{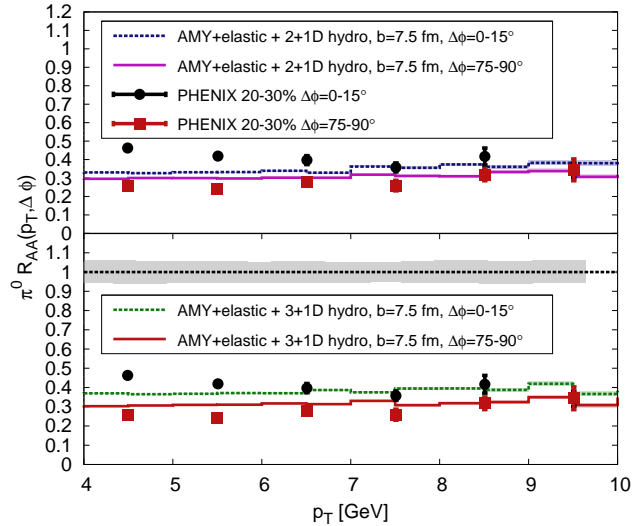}
    \caption{(Color online) Nuclear modification factor as a function of $p_T$ and $\Delta \phi$ for mid-central 
      collisions at $\sqrt{s}=200\,{\rm GeV}$.
      MARTINI results ($b=7.5\,{\rm fm}$, upper panel 2+1D hydro, $\alpha_{\rm s}=0.33$; lower panel 3+1D hydro, $\alpha_{\rm s}=0.3$) 
      with both radiative and elastic processes for most in-plane
      and most out-of-plane angular bins compared to PHENIX data from \cite{Afanasiev:2009iv}. The gray band indicates the experimental 
      error in $R_{AA}$.}
    \label{fig:raaphi}
  \end{center}
\end{figure}

In Figs.~\ref{fig:raa} and \ref{fig:kolbraa} the neutral pion $R_{AA}$ at midrapidity is averaged over the azimuthal angle with 
respect to the reaction plane $\phi$. 
To learn more about the produced medium, e.g. to disentangle the effects of the density of the medium and the pathlength 
traversed on the energy loss, $R_{AA}$ has been studied at midrapidity in {\em non-central} collisions not only as a function of
$p_T$ but also as a function of the azimuth $\phi$ in \cite{Adler:2006bw,Wei:2009mj,Afanasiev:2009iv}.
We perform the same analysis within MARTINI and separate the azimuth in 6 bins of $15^{\circ}$ each, reaching from $0-15^{\circ}$ (in-plane)
to $75-90^{\circ}$ (out-of-plane), as shown in Fig.~\ref{fig:anglephi}.
$R_{AA}$ is then determined in each bin separately. 
We employ both 2+1 and 3+1 dimensional hydrodynamic backgrounds and $\alpha_{\rm s}=0.33$ 
and $\alpha_{\rm s}=0.3$, respectively.
Fig.~\ref{fig:raaphi} shows the most extreme cases 
($0-15^{\circ}$ and $75-90^{\circ}$) at $b=7.5\,{\rm fm}$ compared to recent mid-central (20-30\%) data by PHENIX \cite{Afanasiev:2009iv}.

Particularly at lower $p_T$, we find a less pronounced difference between the in-plane and out-of-plane 
result than the data, as do previous theoretical calculations \cite{Qin:2007zzf,Wei:2009mj}.
Furthermore, one can see that the result is closer to the experimental data for the 3+1 dimensional hydrodynamic background. 
This is due to a larger initial eccentricity in this case.
Generally, we do not expect to describe the low $p_T$ part correctly, because we
do not incorporate soft physics such as flow effects or recombination.
The too small difference for $p_T\gtrsim 4\,{\rm GeV}$ found in the simulation could in part stem from 
a too small eccentricity of the initial state when using Glauber initial conditions in the hydrodynamic
calculation, or a too large surface bias in the theoretical description.
Use of Color-Glass-Condensate initial conditions in 3+1 dimensional hydrodynamics may help to improve the desciption of the data 
\cite{Hirano:2005xf,Drescher:2006pi}.


\subsection{Photon production}

Photons in the high $p_T$ region produced in nuclear collisions are dominantly
direct photons, fragmentation photons, and jet-plasma photons. 
Direct photons are included in PYTHIA.
Apart from leading order direct photons, PYTHIA produces additional photons emitted during the vacuum showers. 
Parts of these overlap with effects that are found in NLO calculations \cite{Aurenche:1986ff,Aurenche:1987fs}, 
but there is no simple theoretical way to identify the amount of overlap between the two.
Also fragmentation photons are part of the photons produced in the showers in PYTHIA. 
We computed photon production within PYTHIA with and without photons from showers and found 
that the shower contribution leads to an effective $K$-factor of approximately 1.8 in the regarded $p_T$ range.

For heavy-ion reactions MARTINI adds the very relevant jet-medium photons from photon radiation 
$q\rightarrow q\gamma$ (bremsstrahlung photons) and jet-photon conversion via 
Compton and annihilation processes.
As above, we include the full vacuum shower in the shown calculation - most of the shower photons will be emitted
before the medium has formed and the parton has realized that it has formed. To improve on this approximation and include
interference between vacuum and in-medium radiation is the subject of future work.
  
\begin{figure}[tb]
  \begin{center}
    \includegraphics[width=8.5cm]{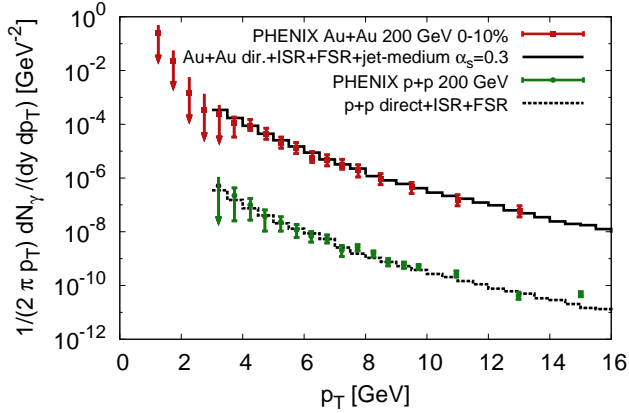}
    \caption{(Color online) Photon yield in p+p and Au+Au collisions at RHIC energy $\sqrt{s}=200\,{\rm GeV}$.
    MARTINI results ($b=2.4\,{\rm fm}$, 3+1D hydro) compared to data from \cite{Adler:2006yt,Adare:2008fqa,Isobe:2007ku}. }
    \label{fig:photons}
  \end{center}
\end{figure}

We present results for photon production in p+p and Au+Au collisions at $\sqrt{s}=200\,{\rm GeV}$ 
compared to data by PHENIX  \cite{Adler:2006yt,Adare:2008fqa,Isobe:2007ku} in Fig.~\ref{fig:photons}.

\begin{figure}[tb]
  \begin{center}
    \includegraphics[width=8.5cm]{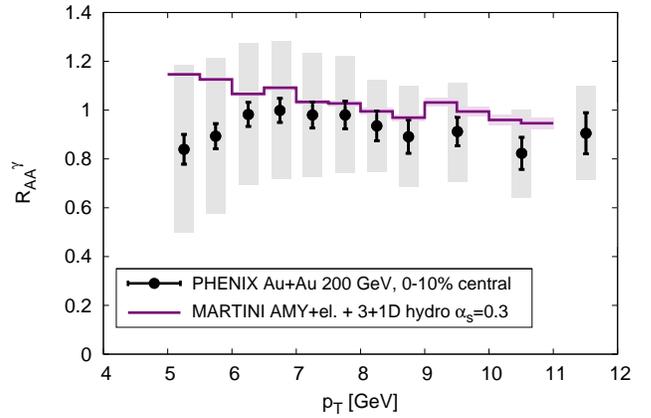}
    \caption{(Color online) Photon nuclear modification factor for central collisions at $\sqrt{s}=200\,{\rm GeV}$.
    MARTINI results compared to data from \cite{Isobe:2007ku}. See main text for details.}
    \label{fig:photon-raa}
  \end{center}
\end{figure}

Another observable that provides information on the effect of the nuclear medium on 
photon production is the photon nuclear modification factor
\begin{equation}
  R_{AA}^\gamma=\frac{1}{N_{\rm coll}(b)} \frac{dN_{AA}^\gamma(b)/d^2p_T dy}{dN_{pp}^\gamma/d^2p_Tdy}\,.
\end{equation}
Fig.~\ref{fig:photon-raa} shows $R_{AA}^\gamma$ as a function of $p_T$ for most central Au+Au collisions 
($b=2.4\,{\rm fm}$) at RHIC compared with $0-10\%$ central PHENIX data. 
We find that in the approximation of 
including all photons from the vacuum shower leads to good agreement with the data within the error bars.
We checked that not including any photons from the final state vacuum shower before $\tau_0$, which correspond to a pre-equilibium 
contribution, reduces $R_{AA}$ by approximately $20\%$. 



\section{Conclusions and Outlook}
\label{conclusions}
We presented first results obtained with the newly developed
Modular Algorithm for Relativistic Treatment of heavy IoN Interactions (MARTINI).
This hybrid approach describes the soft background medium using hydrodynamics or other medium models
and simulates the hard event microscopically, using PYTHIA~8.1 to generate individual hard
nucleon-nucleon collisions. Hard partons are evolved through the medium using the McGill-AMY
evolution scheme including radiative and elastic processes.
Fragmentation is performed employing PYTHIA~8.1 which uses the Lund string fragmentation model. 
Apart from parameters in PYTHIA which were fixed by matching the neutral pion and photon spectra in p+p collisons
to experimental data, $\alpha_{\rm s}$ is the only free parameter.
Employing the 3+1 dimensional hydrodynamic evolution from \cite{Nonaka:2006yn}, it was set to
$\alpha_s=0.3$ to match the neutral pion $R_{AA}$ measurement for central collisions.
Using the same value for all other calculations (there was no additional freedom in any of the calculations), 
we were able to describe the neutral pion $R_{AA}$ in mid-central collisions as well as photon yields and photon $R_{AA}$. 
For photon production, we added photons from jet-medium interactions to the prompt and shower photons  
that PYTHIA produces.

We have demonstrated that MARTINI is able to reveal the effect of using
different evolution models for the soft background on results for hard observables.
For example, when using a 2+1 dimensional hydrodynamical evolution model \cite{Kolb:2000sd,Kolb:2002ve,Kolb:2003dz}, 
a slightly different value for $\alpha_{\rm s}$ ($\alpha_{\rm s}=0.33$) had to be employed to describe the experimental data.
The calculated  azimuthal dependence of the neutral pion $R_{AA}$ in mid-central collisions turned out to be weaker
than that found by PHENIX, for both shown hydrodynamic models, but more so for the 2+1 dimensional one.
This is mainly due to a larger initial eccentricity in the 3+1 dimensional model.
The result for the 3+1 dimensional hydrodynamic background 
is along the line with previous calculations using different energy loss schemes
\cite{Qin:2007zzf,Wei:2009mj}. 
The weak azimuthal dependence, particularly for $p_T\lesssim 6\,{\rm GeV}$,
can be due to the lack of soft physics effects such as flow and recombination, as well as
a too small eccentricity of the initial state when using Glauber initial conditions for the hydro evolution calculation.

Having shown that MARTINI can reproduce one-body observables in good agreement 
with the data, the next step will be to explore its full potential 
by studying many-body observables and correlations.
Another future task is the implementation of heavy quark evolution.

\section*{Acknowledgments}
\hyphenation{Abhijit Majumder Ulrich Heinz Michael Strickland Richard Tomlinson Steffen Bass Evan Frodermann
Chiho Nonaka Huichao Song Guy Moore Guang Qin Vasile Topor-Pop}
We are happy to thank Steffen Bass, Evan Frodermann, Ulrich Heinz, Scott Moreland,
Chiho Nonaka, and Huichao Song for very useful correspondence. 
We thank Torbj\"orn Sj\"ostrand for very helpful clarifications regarding PYTHIA~8.1, and gratefully acknowledge fruitful
discussions with Guy Moore, Guang-You Qin, Thorsten Renk, Michael Strickland, and Vasile Topor-Pop.
This work was supported in part by the Natural Sciences and Engineering Research Council of Canada. 
B.S.\ gratefully acknowledges a Richard H.~Tomlinson Fellowship awarded by McGill University.
\bibliography{martini}

\end{document}